\pdfoutput=1
\documentclass[runningheads]{llncs}
\usepackage[T1]{fontenc}
\usepackage{graphicx}
\usepackage{amssymb}
\usepackage{amsmath}
\usepackage{multirow}
\usepackage{booktabs}
\usepackage{bbding}
\usepackage{subcaption}
\usepackage[colorlinks,
            linkcolor=blue,
            anchorcolor=blue,
            citecolor=blue,
            urlcolor=blue
            ]{hyperref}

\begin{document}
\title{SkinDualGen: Prompt-Driven Diffusion for Simultaneous Image-Mask Generation in Skin Lesions}
%
%
\author{Zhaobin Xu\inst{1 (}\Envelope\inst{)}\orcidID{0009-0004-3517-7911}}
\authorrunning{F. Author et al.}
%
\institute{Shandong University, Qingdao, China \\
\email{sea.xuo@gmail.com}\\}
\maketitle              
\begin{abstract}
Medical image analysis plays a pivotal role in the early diagnosis of diseases such as skin lesions. However, the scarcity of data and the class imbalance significantly hinder the performance of deep learning models. We propose a novel method that leverages the pretrained Stable Diffusion-2.0 model to generate high-quality synthetic skin lesion images and corresponding segmentation masks. This approach augments training datasets for classification and segmentation tasks. We adapt Stable Diffusion-2.0 through domain-specific Low-Rank Adaptation (LoRA) fine-tuning and joint optimization of multi-objective loss functions, enabling the model to simultaneously generate clinically relevant images and segmentation masks conditioned on textual descriptions in a single step. Experimental results show that the generated images, validated by FID scores, closely resemble real images in quality. A hybrid dataset combining real and synthetic data markedly enhances the performance of classification and segmentation models, achieving substantial improvements in accuracy and F1-score of 8\% to 15\%, with additional positive gains in other key metrics such as the Dice coefficient and IoU. Our approach offers a scalable solution to address the challenges of medical imaging data, contributing to improved accuracy and reliability in diagnosing rare diseases.
\end{abstract}
\section{Introduction}
Medical image analysis serves as a cornerstone of modern medicine, particularly in the early detection of skin lesions. Despite its importance, medical image analysis faces two primary challenges: the scarcity of high-quality annotated data and class imbalance. Firstly, acquiring medical images demands specialized equipment and personnel, while the annotation process relies on experienced dermatologists, rendering data collection both costly and time-intensive. Secondly, skin lesion datasets frequently exhibit severe class imbalance, with benign lesions (e.g., nevi) vastly outnumbering malignant ones (e.g., melanoma). This disparity biases deep learning models toward majority classes during training, compromising their ability to detect rare, minority-class lesions accurately.

Conventional data augmentation techniques—such as rotation, flipping, and color adjustments—offer limited diversity and fail to adequately address class imbalance, despite marginally increasing data volume~\cite{shorten2019survey}. Consequently, synthetic data generation has emerged as a promising approach to overcome these limitations by producing diverse, high-quality images and annotations. Koetzier et al.~\cite{koetzier2024generating} reviewed the utility of generative models in medical imaging, while Ktena et al.~\cite{ktena2024generative} demonstrated that generative models can enhance classifier fairness, providing novel strategies to mitigate class imbalance. In recent years, Stable Diffusion, a pretrained diffusion-based model, effectively generates high-quality images from text prompts, providing robust flexibility and control~\cite{rombach2022high}.

This paper introduces a method that utilizes Stable Diffusion to generate synthetic skin lesion images and segmentation masks, thereby enriching training datasets for classification and segmentation tasks. We fine-tune Stable Diffusion using LoRA~\cite{hu2022lora}, propose a one-prompt dual-generation technique. Our primary contributions are threefold:
\begin{enumerate}
    \item \textbf{Domain-Adaptation Fine-Tuning:} We adapt Stable Diffusion-2.0 to the medical imaging domain through LoRA fine-tuning and joint optimization of multi-objective loss functions, enabling the generation of high-quality synthetic skin lesion image-mask pairs;
    \item \textbf{One-Prompt Dual-Generation:} We develop a technique to simultaneously generate image-segmentation mask pairs from a single prompt, ensuring efficiency and consistency in the output;
    \item \textbf{Performance Validation:} We verify performance gains in classification and segmentation tasks using a hybrid dataset of real and synthetic data, evaluated on the ISIC-GPT and HAM10000 datasets.
\end{enumerate}
For this paper, the code is public at \href{https://github.com/JaspinXu/SkinDualGen}{here}.

\section{Related  Works}
\subsection{Medical Images Generation}
Medical image generation techniques address the challenges of data scarcity and privacy constraints by producing realistic synthetic images.

Variational Autoencoders (VAEs) generate new samples by learning a probabilistic mapping from images to a latent space, optimized through minimizing reconstruction error and Kullback-Leibler (KL) divergence~\cite{kingma2013auto}. In medical imaging, VAEs are widely used for reconstruction, anomaly detection, and data augmentation. For example, Baur et al. applied VAEs to generate brain MRI images, improving tumor detection accuracy~\cite{baur2021autoencoders}. However, VAE-generated images often lack sharpness and high-frequency details, which can hinder their applicability in tasks requiring precise anatomical delineation.

Generative Adversarial Networks (GANs) produce high-fidelity images via adversarial training between a generator and a discriminator~\cite{goodfellow2014generative}. In skin lesion segmentation, Innani et al.'s EGAN framework achieved precise results through unsupervised adversarial learning, though it demanded high-quality generators and discriminators, increasing computational complexity~\cite{innani2023generative}. Despite some strengths, GANs frequently encounter mode collapse and training instability, reducing image diversity and clinical utility.

Diffusion models have recently emerged as a robust alternative to GANs, generating high-quality images through iterative denoising processes. Ho et al.'s Denoising Diffusion Probabilistic Model (DDPM) laid the theoretical groundwork~\cite{ho2020denoising}. And Stable-Diffusion is increasingly adopted. Kazerouni et al. provided an extensive review of diffusion models in medical imaging, highlighting their strengths in synthesis, reconstruction, and augmentation, as well as their ability to enhance downstream task performance~\cite{kazerouni2023diffusion}. In dermatology, Bozorgpour et al.'s DermoSegDiff model leveraged boundary-aware diffusion to achieve high-precision segmentation, surpassing GAN-based methods~\cite{bozorgpour2023dermosegdiff} while Yu et al.'s MedDiff-FM exploited diffusion models' versatility across various medical imaging tasks~\cite{yu2024meddiff}.

In contrast to prior diffusion-based research, our study introduces a novel approach by generating both images and segmentation masks from a single prompt, streamlining the process. Additionally, we integrate large language models to enrich clinical descriptions, thereby improving data diversity and relevance for medical applications.
\subsection{Data Augmentation}
Data augmentation enhances the robustness and generalization of medical image analysis models. Traditional techniques, such as geometric transformations (e.g., rotation, flipping, scaling) and color adjustments (e.g., brightness, contrast), are simple to implement but provide limited diversity. Chlap et al. reviewed their use in medical imaging, noting their insufficiency for capturing complex lesion morphologies~\cite{chlap2021review}. Generative models offer a solution by producing diverse, clinically relevant samples. Abdelhalim et al.'s self-attention progressive GAN (PGAN) augmented skin lesion datasets, boosting classification performance~\cite{abdelhalim2021data}. Diffusion models exhibit even greater potential. Montoya et al.'s MAM-E model generated mammograms for breast cancer classification, outperforming traditional augmentation techniques~\cite{montoya2024mam}.

Our method could capitalize on Stable Diffusion’s conditional generation capabilities to target minority classes, addressing class imbalance effectively. Through fine-tuning and conditional prompts, we ensure that synthetic images closely align with real images in clinically significant features, such as texture and boundaries, enhancing their utility in medical image analysis.

\section{Method}
\begin{figure}[!htbp]
\vspace{-2.0em}
\setlength{\floatsep}{5pt plus 2pt minus 2pt}
\setlength{\textfloatsep}{5pt plus 2pt minus 2pt}
\setlength{\intextsep}{5pt plus 2pt minus 2pt}
\includegraphics[width=\textwidth]{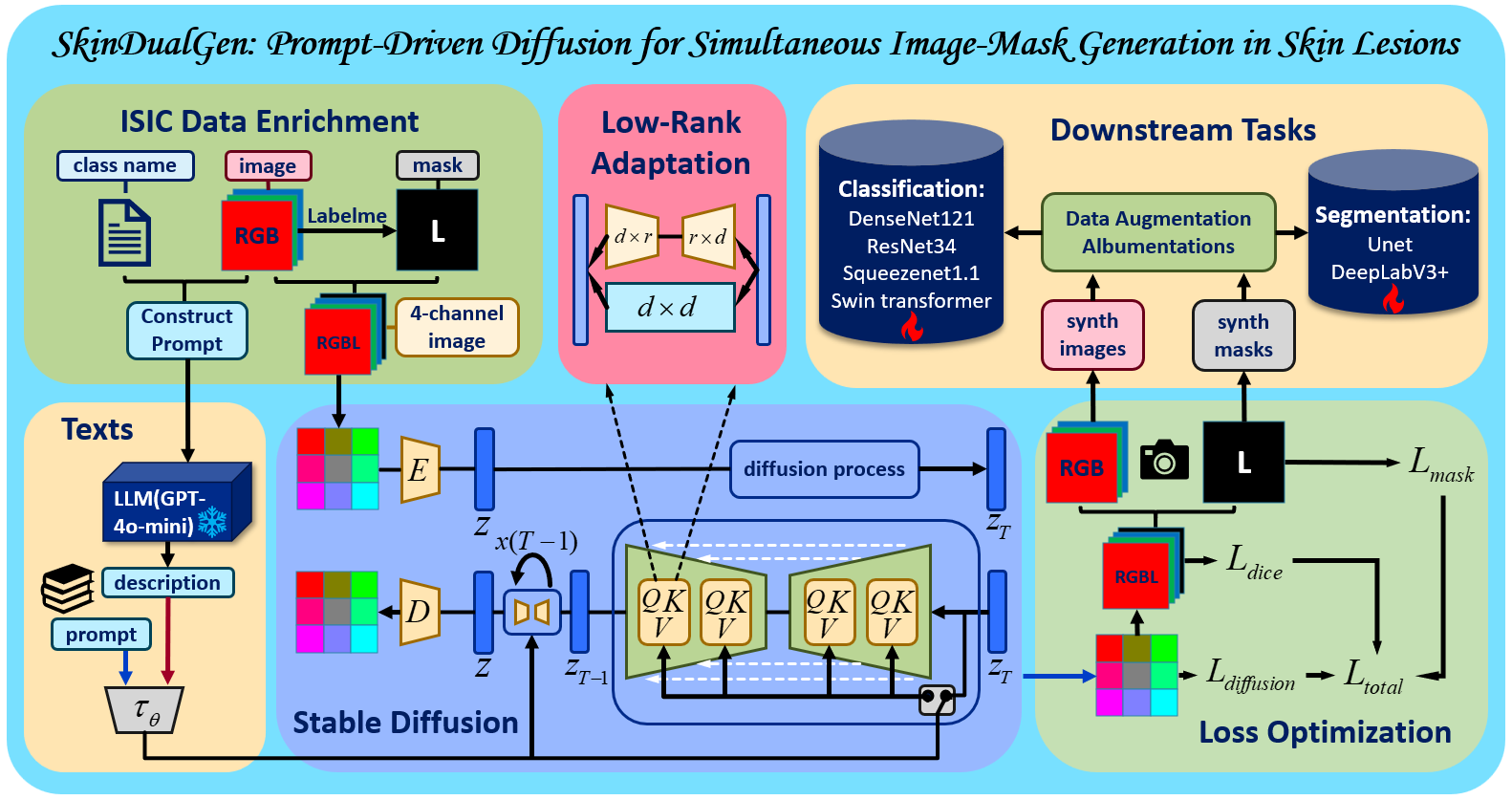}
\caption{The pipeline of our proposed SkinDualGen approach} \label{fig1}
\end{figure}

Our method aims to generate high-quality synthetic skin lesion images and their segmentation masks using Stable Diffusion to augment the training dataset for classification and segmentation tasks.The detailed pipeline is shown in Fig.~\ref{fig1}.

\subsection{Stable Diffusion}
Diffusion models represent a class of powerful generative models capable of producing high-quality images through an iterative denoising process. The Denoising Diffusion Probabilistic Model (DDPM), proposed by Ho et al.~\cite{ho2020denoising}, established the theoretical foundation for diffusion models. The forward diffusion process incrementally introduces Gaussian noise to the data across multiple timesteps, mathematically expressed as:
\begin{equation}
q(x_t|x_{t-1})=\mathcal{N}(x_t;\sqrt{1-\beta_{t}}x_{t-1},\beta_tI)
\end{equation}
where $\beta_t$ denotes the noise schedule at timestep $t$, and $x_t$ represents the noisy data. The reverse process learns to recover the original data by modeling the denoising distribution:
\begin{equation}
p_\theta(x_{t-1}|x_t)=\mathcal{N}(x_{t-1};\mu_\theta(x_t,t),\sigma_\theta(x_t,t)I)
\end{equation}
where $\mu_\theta$ and $\sigma_\theta$ are parameters predicted by a UNet architecture.

Stable Diffusion, introduced by Rombach et al.~\cite{rombach2022high}, enhances computational efficiency by conducting the diffusion process within the latent space of a VAE. The workflow involves encoding an image into a latent representation , performing diffusion and denoising in this latent space, and decoding the resulting latent back into the image domain. This generation process can be formulated as:
\begin{equation}
z_T\sim\mathcal{N}(0,1),z_0=f_\theta(z_T,c),x=Decoder(z_0)
\end{equation}
where $z_T$ is the initial noise, $c$ is the conditioning input (e.g., a text prompt), and $f_\theta$ denotes the denoising network (UNet).

Owing to their ability to generate high-quality images, diffusion models have gained prominence in medical imaging. In this study, we adopt Stable Diffusion-2.0 as our baseline model due to its optimal balance of performance and compatibility with consumer-grade hardware.

\subsection{Low-Rank Adaptation}
To efficiently tailor Stable Diffusion to the medical imaging domain, we employ Low-Rank Adaptation (LoRA)~\cite{hu2022lora}. LoRA approximates weight updates using low-rank matrices, substantially reducing computational overhead while maintaining model efficacy. For a pre-trained weight matrix $W\in\mathbb{R}^{d\times k}$, LoRA expresses the weight update as:
\begin{equation}
W+\Delta W,\Delta W=AB
\end{equation}
where $A\in\mathbb{R}^{d\times r}$, $B\in\mathbb{R}^{r\times k}$, and $r\ll \min{(d,k)}$ is the rank, Only  and are trainable, significantly decreasing the number of parameters requiring optimization.

We integrate LoRA adapters into the attention layers of Stable Diffusion’s UNet, which are critical for incorporating text-based conditions to align generated images with clinical descriptions. This approach reduces the number of trainable parameters by over 90\%. Furthermore, LoRA is extensible and can be combined with other techniques such as ControlNet for additional conditional control~\cite{zhang2023adding}, or DreamBooth for personalized generation~\cite{ruiz2023dreambooth}.

\subsection{One-Prompt-Dual-Generation}
To simultaneously generate RGB images and segmentation masks from a single prompt, we construct four-channel images, with the first three channels representing the RGB image and the fourth channel encoding the segmentation mask. We modify the VAE architecture by adjusting the encoder’s input convolutional layer to accept four-channel inputs (RGB + mask) and the decoder’s output layer to produce four-channel outputs. Additionally, the UNet’s first convolutional layer is updated to process four-channel latent inputs. In a nutshell, this unified four-channel representation (RGB + mask) enables the VAE to jointly encode and decode images and masks, streamlining the generation process while maintaining consistency.

The training process leverages multi-task learning, optimizing a combination of loss functions. The primary diffusion loss is defined as the mean squared error (MSE) between the predicted and actual noise in the latent space:
\begin{equation}
L_{diffusion}=E_{t,z_t,\epsilon}\left[\parallel \epsilon-\epsilon_\theta(z_t,t,c)\parallel_2^2\right]
\end{equation}

We also incorporate a binary cross-entropy (BCE) loss to assess the discrepancy between the predicted mask logits and the ground truth:
\begin{equation}
L_{mask}=-\frac{1}{N}\sum_{i=1}^N\left[y_i\log{(\hat{y_i})}+(1-y_i)\log{(1-\hat{y_i})}\right]
\end{equation}

Additionally, we employ the Dice loss to enhance the overlap between predicted and actual mask regions:
\begin{equation}
L_{dice}=1-\frac{2\sum_{i=1}^Ny_i\hat{y_i}+\epsilon}{\sum_{i=1}^Ny_i+\sum_{i=1}^N\hat{y_i}+\epsilon}
\end{equation}
where $\epsilon$ is a smoothing term to ensure numerical stability.
The total loss is a weighted combination of these components:
\begin{equation}
L_{total}=\lambda_1L_{diffusion}+\lambda_2L_{mask}+\lambda_3L_{dice}
\end{equation}

During training, the four-channel images are encoded into the VAE’s latent space, subjected to noise at random timesteps, and denoised by the UNet. In inference, a single prompt produces a four-channel output. The novelty of this approach lies in its efficiency and consistency.

\subsection{Datasets Enrichment based on Large Language Model}
\label{sec1}
In the field of dermatology, publicly available datasets are limited, with the \href{https://challenge.isic-archive.com/data/}{ISIC} series being among the most prominent, encompassing over 13,000 dermoscopic images. These datasets, curated by the International Skin Imaging Collaboration (ISIC). Most images are accompanied by clinical metadata, which have been reviewed and annotated by recognized experts, including detailed dermoscopic features. However, the sheer volume of these datasets presents challenges for individual researchers conducting large-scale analyses.

For this study, we selected the ISIC 2020 dataset due to its diverse range of skin lesion categories. From this dataset, we curated a refined subset comprising 1,990 images across seven categories. As the original ISIC 2020 dataset lacks segmentation masks, we employed the Labelme tool to manually annotate each image, generating corresponding segmentation masks to support downstream segmentation tasks.

In recent years, Large Language Models (LLMs) have emerged as powerful tools in medical imaging, owing to their exceptional capabilities in text comprehension, analysis, and generation. Our approach involved retrieving each category along with its corresponding RGB skin lesion images and using a meticulously designed prompt to guide GPT-4o-mini in generating detailed descriptions. These descriptions aim to encapsulate fine-grained features using professional medical terminology, adhering to the prompt structure: "\textit{Analyze this {category} dermatology image. Describe in medical terms and give a sentence. Use ICD-11 terminology and begin with 'a dermoscopic lesion photo of {category} for skin cancer diagnosis,...'}". Unlike Medghalchi et al.’s approach of merely employing diverse adjectives to expand monotonous fixed statements~\cite{medghalchi2024meddap}, our LLM-based semantic expansion ensures greater diversity and specificity. Following this process, we generated detailed descriptions for each image, reviewed by dermatologists. The resulting dataset is well-suited for small-scale multimodal research prioritizing richness.

\subsection{Synthesize Images and Masks for Downstream Tasks}
The high-quality synthetic images and masks generated in this study are utilized to augment training datasets for classification and segmentation tasks, with the goal of enhancing model performance and mitigating class imbalance issues.

In classification tasks, class imbalance poses a significant challenge, with certain lesion types (e.g., melanoma) being underrepresented. To address this, we could generate additional synthetic samples for minority classes using text prompts such as "\textit{a dermoscopic lesion photo of melanoma for skin cancer diagnosis}" to produce melanoma image-mask pairs. These synthetic images were combined with real images, significantly improved model generalization by increasing data diversity. For segmentation tasks, synthetic masks provide additional training pairs, enabling the model to learn more lesion boundaries. The consistency between synthetic masks and their corresponding images ensures that the model can effectively learn from diverse samples. During training, both real and synthetic image-mask pairs were employed, expanding the dataset and reducing the risk of overfitting.

In addition to synthetic data, we incorporated traditional data augmentation techniques to further enrich the dataset. For classification tasks, we applied transformations from the Torchvision library, while for segmentation tasks, we utilized the Albumentations library, ensuring consistency between images and masks, markedly enhancing the performance of downstream tasks.

\section{Experiments}
\subsection{Experimental Setting}
\subsubsection{Datasets}
\paragraph{Skin Cancer ISIC and GPT-based Descriptions:}
As outlined in Sec.~\ref{sec1}, this novel dataset comprises 1,990 skin lesion images spanning seven categories and consists of three components: RGB images, L masks, and JSON descriptions. It is publicly accessible on \href{https://www.kaggle.com/datasets/zhaobinxu/skin-cancer-isic-and-gpt-based-descriptions}{Kaggle}. The training set was employed to fine-tune Stable Diffusion-2.0 and constitutes the real data portion of the hybrid training set for downstream classification and segmentation models. The test set was used to evaluate the performance of the trained downstream models.
\paragraph{HAM10000:}
The HAM10000 (Human Against Machine with 10,000 training images) dataset~\cite{tschandl2018ham10000} is a comprehensive collection of 10,015 color images designed for the study and training of skin lesion classification models. The dataset exhibits significant class imbalance, presenting challenges for model training. In this study, it serves as a test set to evaluate the generalizability of classification and segmentation models trained on the hybrid dataset to unseen data.

\subsubsection{Models}
\paragraph{Image-Mask Generation:}
For the generation of images and masks, we selected SD2.0 and fine-tuned it using LoRA~\cite{hu2022lora}. Among the various open-source models in the Stable Diffusion series, SD1.4 and SD1.5 employ a general CLIP model, which demonstrates suboptimal text-image alignment. SD2.1 utilizes OpenCLIP, offering improved performance; however, it requires input and output resolutions of 768 pixels and occasionally generates crow-scaled noise. SDXL and SD3.5, due to their extensive parameter counts, are incompatible with consumer-grade GPUs. Thus, we opted for SD2.0 as the base model for modification and fine-tuning, balancing performance and hardware compatibility.

\paragraph{Classification:}
For classification tasks, we utilized four models: DenseNet121, ResNet34, SqueezeNet1.1, and Swin Transformer. DenseNet121~\cite{huang2017densely}~features dense connections where each layer is directly linked to all preceding layers. ResNet34~\cite{he2016deep}~incorporates residual connections that add the input directly to the output, facilitating the training of deep networks with stable and robust performance. SqueezeNet1.1~\cite{iandola2016squeezenet}, a lightweight network, significantly reduces parameters while preserving performance. Swin Transformer~\cite{liu2021swin}~employs a hierarchical design and captures long-range dependencies.

\paragraph{Segmentation:}
For segmentation tasks, we adopted U-Net and DeepLabV3+. U-Net \cite{ronneberger2015u} features an encoder-decoder structure where the encoder extracts image features and the decoder restores spatial information. DeepLabV3+~\cite{chen2018encoder}~combines convolutional neural networks and Atrous Spatial Pyramid Pooling (ASPP). The encoder employs atrous convolution to expand the receptive field and capture multi-scale information, while the decoder upsamples and fuses low-level features to recover details, achieving high precision in complex scenes and high-resolution image segmentation.

\subsubsection{Metrics}
\paragraph{Image-Mask Generation:}
To evaluate the similarity between generated and real images, we adopt three widely recognized metrics: Fréchet Inception Distance (FID), Learned Perceptual Image Patch Similarity (LPIPS), and Multi-Scale Structural Similarity (MS-SSIM). FID~\cite{heusel2017gans} assesses the quality of generated images by measuring the distributional similarity between generated and real images within a feature space. A lower FID score indicates a closer match between the distributions, reflecting higher image quality. LPIPS quantifies perceptual similarity by computing a weighted distance between features extracted from pre-trained networks, simulating human visual perception. MS-SSIM evaluates structural consistency by comparing luminance, contrast, and structural information.

\paragraph{Classification:}
For classification tasks, we employ four standard metrics: Accuracy, Sensitivity, Precision, and F1-score. Their formulations are as follows:
\begin{equation}
Accuracy=\frac{TP+TN}{TP+TN+FP+FN},~~Sensitivity=\frac{TP}{TP+FN}
\end{equation}
\begin{equation}
Precision=\frac{TP}{TP+FP},~~F1score=2\times\frac{Precision\times Sensitivity}{Precision+Sensitivity}
\end{equation}
where $TP$ represents true positives, $TN$ true negatives, $FP$ false positives, and $FN$ false negatives.
\paragraph{Segmentation:}
For segmentation tasks, we utilize four established metrics: Dice Coefficient, Intersection over Union (IoU), Average Surface Distance (ASD), and Hausdorff Distance (HD). The Dice Coefficient~\cite{dice1945measures} measures the overlap between predicted and ground truth segmentation regions. Ranging from 0 to 1, higher values indicate better segmentation accuracy. IoU~\cite{everingham2015pascal}, another overlap-based metric, is commonly used in segmentation evaluation, with values approaching 1 reflecting superior performance. ASD calculates the average distance between the boundaries of predicted and ground truth segmentations; smaller values indicate closer alignment. HD quantifies the maximum distance between predicted and ground truth boundaries; lower values are preferable. Their definitions are as follows:
\begin{equation}
Dice=\frac{2|\hat{m}\cap m|}{|\hat{m}|+|m|},~~IoU=\frac{|\hat{m}\cap m|}{|\hat{m}\cup m|}
\end{equation}
\begin{equation}
ASD=\frac{1}{2}\left(\frac{1}{|\hat{m}|}\sum_{p\in\hat{m}}d(p,m)+\frac{1}{|m|}\sum_{p\in m}d(p,\hat{m})\right)
\end{equation}
\begin{equation}
HD=\max\left(\max_{p\in\hat{m}}d(p,m),\max_{p\in m}d(p,\hat{m})\right)
\end{equation}
where $\hat{m}$ and $m$ denote the predicted and ground truth segmentation regions, respectively, and $d(\cdot,\cdot)$ represents the minimum distance from a point to a set. Dice and IoU emphasize region overlap, while ASD and HD focus on boundary accuracy, with ASD reflecting average error and HD capturing maximum deviation.

\subsection{Fine-tuning and Making Expanded Datasets with SD2}
This section outlines the experimental setup leveraging the pretrained Stable Diffusion-2.0 checkpoint, a text-to-image generative model sourced from the \href{https://huggingface.co/stabilityai/stable-diffusion-2}{Hugging Face}, chosen for its robust generative capabilities.

For fine-tuning, we employed the novel ISIC-GPT dataset introduced in the previous section. The Stable Diffusion-2.0 model was adapted to process four-channel inputs and outputs, with LoRA applied to efficiently fine-tune the attention layers of the UNet component. After evaluating ranks of 2, 4, and 8, we selected a rank of 4 for LoRA fine-tuning of the attention layers, as it achieved an optimal balance between image fidelity and structural consistency. $\lambda$ is simply set to 1. The base UNet and VAE parameters remained frozen, with only the LoRA parameters trained, substantially reducing the number of trainable parameters, computational cost, and training duration. RGB images were resized to 256×256 using LANCZOS interpolation, while L-format segmentation masks were resized using nearest-neighbor interpolation. Data augmentation was conducted using the Albumentations, applying pixel-level transformations and spatial transformations, synchronized across RGB images and masks. During data loading, a batch size of 4 was used with 12 worker processes to accelerate retrieval. Fine-tuning spanned 100 epochs with a learning rate of 1e-4, a batch size of 4, and the AdamW optimizer was employed for efficiency. Training was performed on an NVIDIA GeForce RTX 4090 GPU, completing in approximately 5 hours.

\begin{figure}[!htbp]
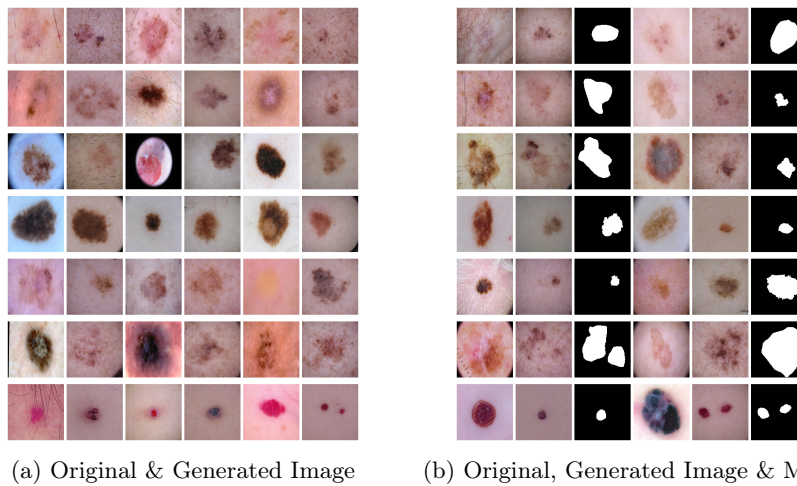

\vspace{-2.0em}
\setlength{\floatsep}{5pt plus 2pt minus 2pt}
\setlength{\textfloatsep}{5pt plus 2pt minus 2pt}
\setlength{\intextsep}{5pt plus 2pt minus 2pt}
    \centering
    \begin{subfigure}{0.48\linewidth}
        \centering
        \includegraphics[height=0.3\textheight]{comparison1.pdf}
        \caption{Original \& Generated Image}
    \end{subfigure}
    \centering
    \begin{subfigure}{0.48\linewidth}
        \centering
        \includegraphics[height=0.3\textheight]{comparison2.pdf}
        \caption{Original, Generated Image \& Mask}
    \end{subfigure}
    \caption{Comparison of Image-Mask Pairs(One row corresponds to one category)}
    \label{fig3}
\vspace{-2.0em}
\end{figure}

In the inference phase, four-channel outputs were generated and split into RGB and L channels. RGB channels were denormalized to their original range, while the L channel was processed via a Sigmoid function with a 0.5 threshold to yield a binary mask, saved in PNG format. The prompt structure guiding the diffusion model was "\textit{a dermoscopic lesion photo of \{class\_name\} for skin cancer diagnosis,}" where {class\_name} represents one of seven skin lesion categories. Random seeds were used to enhance output diversity, with a resolution of 512×512 pixels and the DDIM scheduler. We used the Optuna optimizer for Bayesian hyperparameter tuning, we identified num\_inference\_steps=45 and guidance\_scale=1.22 as the optimal settings. Image quality assessment for the synthetic dataset, presented in Table~\ref{tab1}, involved computing three metrics between the real training and test sets, and between the real training and synthetic training sets. For a dataset of 1,640 images, an FID below 50 is considered excellent, below 100 acceptable, and above 150 poor. While our synthetic dataset’s FID exceeds that of the real dataset, it remains viable for downstream tasks. The LPIPS and MS-SSIM metrics, though slightly irregular due to the modest dataset size, show comparable values across groups, suggesting sufficient realism in the generated data. Fig.~\ref{fig3} offers a visual comparison of synthetic images and masks against real ones, revealing that synthetic skin lesion images are nearly indistinguishable from real images to the naked eye, underscoring their high fidelity and potential as real-data substitutes.
\begin{table}[!htbp]
\vspace{-2.0em}
\centering
\caption{Image Quality Assessment for Generated Images}\label{tab1}
\begin{tabular}{cccc}
\toprule[1.5pt]
\textbf{Data for calculations} & \textbf{FID} & \textbf{LPIPS} & \textbf{MS-SSIM} \\
\midrule[1pt]
\textbf{Real Train vs Real Test} & 42.058 & 0.545 & 0.190 \\
\textbf{Real Train vs Synth Train} & 68.601 & 0.524 & 0.210 \\
\bottomrule[1.5pt]
\end{tabular}
\vspace{-1.5em}
\end{table}

Consequently, we established three datasets—real, synthetic, and hybrid—all of identical scale, comprising skin lesion images and corresponding masks, primed for subsequent comparative experiments.

\begin{table}[!htbp]
\vspace{-2.0em}
\centering
\caption{Comparison of Performance on Real$\setminus$Synth$\setminus$Hybrid Dataset}\label{tab2}
\resizebox{1.0\linewidth}{!}{
\begin{tabular}{cccccc}
\toprule[1.5pt]
\textbf{Models} & \textbf{Train Dataset} & \textbf{Accuracy} & \textbf{Sensitivity} & \textbf{Precision} & \textbf{F1score} \\
\midrule[1pt]
 & Real only & 71.257 & 71.257 & 73.775 & 71.289 \\
\textbf{DenseNet121} & Synth only & 50.057 & 50.057 & 58.547 & 46.875 \\
& 50\%Real+50\%Synth & \textbf{80.286} & \textbf{80.286} & \textbf{82.160} & \textbf{80.446} \\
\cmidrule[1pt]{2-6}
 & Real only & 71.600 & 71.600 & 74.504 & 71.609 \\
\textbf{ResNet34}& Synth only & 45.943 & 45.943 & 60.142 & 43.055 \\
& 50\%Real+50\%Synth & \textbf{77.257} & \textbf{77.257} & \textbf{79.340} & \textbf{77.454} \\
\cmidrule[1pt]{2-6}
 & Real only & 63.943 & 63.943 & 67.800 & 63.925 \\
\textbf{SqueezeNet1.1} & Synth only & 47.771 & 47.771 & 53.213 & 45.012 \\
& 50\%Real+50\%Synth & \textbf{71.029} & \textbf{71.029} & \textbf{73.747} & \textbf{70.862} \\
\cmidrule[1pt]{2-6}
& Real only & 74.571 & 74.571 & 77.642 & 74.021 \\
\textbf{Swin Transformer} & Synth only & 57.143 & 57.143 & 63.506 & 56.027 \\
& 50\%Real+50\%Synth & \textbf{80.971} & \textbf{80.971} & \textbf{82.690} & \textbf{80.901} \\
\midrule[1.5pt]
\textbf{Models} & \textbf{Train Dataset} & \textbf{Dice} & \textbf{IoU} & \textbf{ASD} & \textbf{HD} \\
\midrule[1pt]
& Real only & 79.397 & 65.836 & \textbf{58.351} & \textbf{95.775} \\
\textbf{U-Net} & Synth only & 64.459 & 47.575 & 64.303 & 114.456 \\
& 50\%Real+50\%Synth & \textbf{80.122} & \textbf{66.840} & 59.101 & 97.882 \\
\cmidrule[1pt]{2-6}
& Real only & 81.373 & 68.595 & 57.063 & 92.113 \\
\textbf{DeepLabV3+} & Synth only & 67.968 & 51.515 & 66.556 & 112.226 \\
& 50\%Real+50\%Synth & \textbf{81.951} & \textbf{69.425} & \textbf{56.023} & \textbf{89.778} \\
\bottomrule[1.5pt]
\end{tabular}}
\vspace{-2.0em}
\end{table}

\subsection{Quantitative and Qualitative Analysis for Tasks}

\subsubsection{Training and Evaluation for Classification Tasks}
In this study, the training data was organized into three configurations: real data only, synthetic data only, and a hybrid dataset comprising 50\% real and 50\% synthetic data. Input images were resized to 224×224 pixels and augmented, followed by normalization. The test set consisted exclusively of real data, resized and normalized without further augmentation. Training was conducted with a batch size of 32, spanning 50 epochs for DenseNet121 and ResNet34, and 40 epochs for SqueezeNet1.1 and Swin Transformer. The learning rate was set to 0.0001. A 5-fold Stratified K-Fold cross-validation approach ensured balanced class representation, with all metrics calculated via macro-averaging.The quantitative results are shown in Table~\ref{tab2}. Overall, the hybrid dataset, integrating the clinical relevance of real data with the diversity of synthetic data, mitigated the distributional constraints of real-only data and the domain gaps of synthetic-only data. This approach enhanced classification performance across all models, with accuracy and F1-score improvements ranging from 8\% to 15\%. Increased data diversity effectively reduced overfitting and improved generalization. The suboptimal results with synthetic-only data likely stem from distributional shifts or label noise relative to the real test set, impeding generalization. Employing diffusion models for data augmentation sharpened inter-class distinctions and reduced intra-class variability, thereby boosting classification accuracy.

\subsubsection{Training and Evaluation for Segmentation Tasks}
For segmentation tasks, we employed U-Net and DeepLabV3+ models, built on ResNet34 and ResNet50 backbones, respectively. The data configurations mirrored those of the classification tasks, with images resized to 512×512 pixels. Training utilized a batch size of 8 over 20 epochs, with a learning rate of 0.001 and Dice Loss. The quantitative results are shown in Table~\ref{tab2}. Overall, the hybrid dataset, by incorporating synthetic mask diversity, refined the models’ capacity to predict edges and fine structures, addressing the shortcomings of real-only data in complex lesion scenarios. This resulted in Dice and IoU improvements of 0.5\%–1\% and substantial reductions in ASD and HD (5–10 units). The poorer performance of synthetic-only data likely arises from geometric distortions in mask generation or domain mismatches with the real test set, leading to imprecise boundary predictions. Dice Loss optimized overlap regions effectively, and mixed-precision training accelerated convergence. DeepLabV3+ consistently outperformed U-Net across all metrics, particularly with the hybrid dataset, achieving a Dice coefficient of 81.951\% and HD of 89.778, highlighting its efficacy in complex medical image segmentation.

\begin{table}[!htbp]
\vspace{-2.0em}
\centering
\caption{Results Evaluating the Robustness on HAM10000}\label{tab3}
\begin{tabular}{cccccc}
\toprule[1.5pt]
\textbf{Models} & \textbf{Dataset} & \textbf{Accuracy} & \textbf{Sensitivity} & \textbf{Precision} & \textbf{F1score} \\
\midrule[1pt]
\multirow{2}{*}{\textbf{DenseNet121}} & Real & 34.930 & 59.227 & 38.660 & 37.633 \\
& Hybrid & \textbf{35.990} & \textbf{66.391} & \textbf{41.249} & \textbf{40.765} \\
\cmidrule[1pt]{2-6}
\multirow{2}{*}{\textbf{ResNet34}}& Real & \textbf{35.968} & 59.645 & 39.001 & 37.898 \\
& Hybrid & 34.333 & \textbf{65.071} & \textbf{39.547} & \textbf{38.655} \\
\cmidrule[1pt]{2-6}
\multirow{2}{*}{\textbf{SqueezeNet1.1}} & Real & 25.368 & 51.408 & 29.833 & 27.495 \\
 & Hybrid & \textbf{27.359} & \textbf{57.198} & \textbf{35.381} & \textbf{33.793} \\
\cmidrule[1pt]{2-6}
\multirow{2}{*}{\textbf{Swin Transformer}}& Real & 34.011 & \textbf{66.194} & 39.908 & 38.197 \\
 & Hybrid & \textbf{34.201} & 60.199 & \textbf{40.169} & \textbf{39.208} \\
\midrule[1.5pt]
\textbf{Models} & \textbf{Dataset} & \multicolumn{2}{c}{\textbf{Dice}} & \multicolumn{2}{c}{\textbf{IoU}} \\
\midrule[1pt]
\multirow{2}{*}{\textbf{U-Net}}& Real &\multicolumn{2}{c}{29.878} & \multicolumn{2}{c}{17.568}\\
 & Hybrid & \multicolumn{2}{c}{\textbf{30.953}} & \multicolumn{2}{c}{\textbf{18.321}} \\
\cmidrule[1pt]{2-6}
\multirow{2}{*}{\textbf{DeepLabV3+}} & Real & \multicolumn{2}{c}{29.399} & \multicolumn{2}{c}{17.235} \\
 & Hybrid & \multicolumn{2}{c}{\textbf{30.110}} & \multicolumn{2}{c}{\textbf{17.730}} \\
\bottomrule[1.5pt]
\end{tabular}
\vspace{-2.0em}
\end{table}

\newpage
\subsubsection{Evaluating the Robustness on HAM10000}

Table~\ref{tab3} presents the robustness evaluation of our models on the HAM10000 dataset, encompassing both classification and segmentation tasks. The hybrid dataset generally enhanced sensitivity and F1-scores in classification tasks, improving positive class detection and overall model performance. However, due to the limited sample size of the training set, accuracy improvements were modest, with some models (e.g., ResNet34) even experiencing slight declines, highlighting the complexity of the HAM10000 dataset. In segmentation tasks, the Dice coefficient and IoU showed limited gains (approximately 1\%), with overall performance remaining low (Dice < 31\%). This may be attributed to constraints in image resolution or annotation quality. Among the models, Swin Transformer and DeepLabV3+ demonstrated relatively superior performance on the hybrid dataset, suggesting potential for further robustness improvements through optimized data preprocessing.
\vspace{-2.0em}
\subsubsection{Visualizing for Classification Tasks}
\begin{figure}[!htbp]
\vspace{-2.0em}
\setlength{\floatsep}{5pt plus 2pt minus 2pt}
\setlength{\textfloatsep}{5pt plus 2pt minus 2pt}
\setlength{\intextsep}{5pt plus 2pt minus 2pt}
\includegraphics[width=\textwidth]{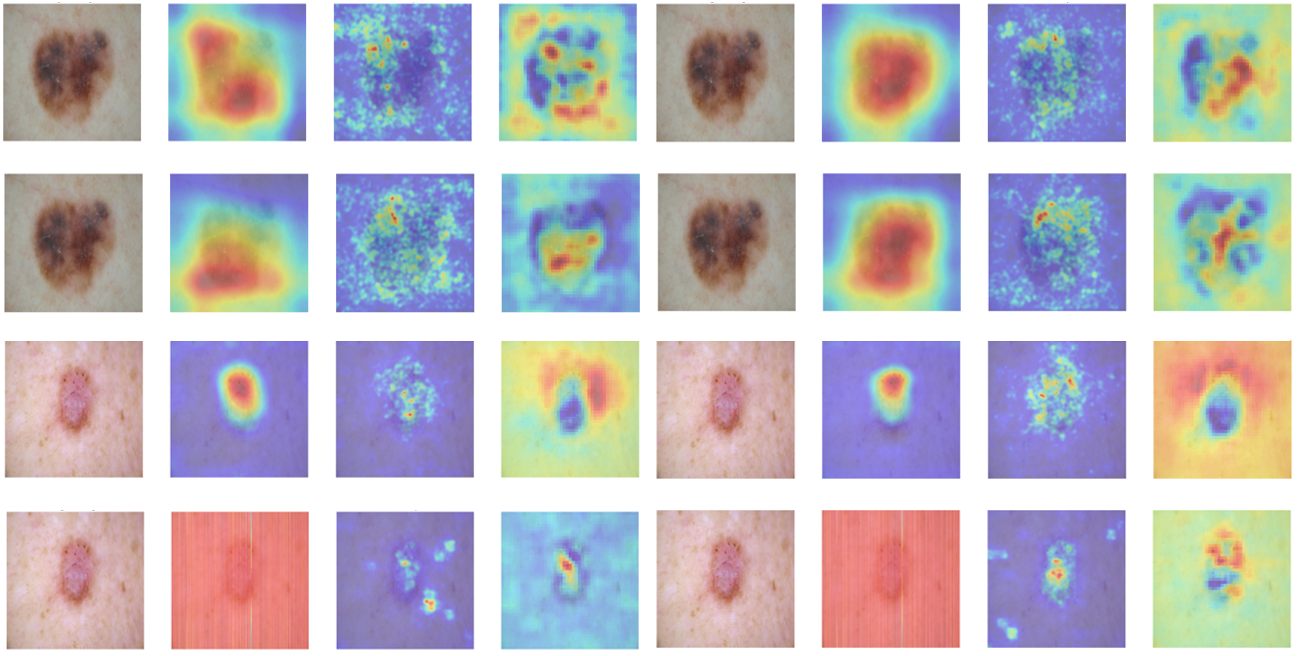}
\caption{Heatmaps of XAI methods (Gradcam, Saliency, Occlusion) for the testset: One row corresponds to one model, the first four images show models trained on the real dataset only, and the next four show models trained on the hybrid dataset. Models trained solely on synthetic data were excluded due to poor performance. Visualized samples are ISIC\_0010034 and ISIC\_0030261, with each set of four images including the original image followed by attention maps from the three XAI methods.} 
\label{fig4}
\vspace{-2.0em}
\end{figure}

Fig.~\ref{fig4} illustrates heatmaps generated using three explainable AI (XAI) techniques—GradCAM, Saliency, and Occlusion—to interpret the decision-making processes of our classification models on the test set. The hybrid dataset enhances the feature space through increased diversity, mitigating overfitting and directing model attention toward pathologically relevant regions. The complementary insights from GradCAM (class-discriminative localization), Saliency (pixel sensitivity), and Occlusion (impact of occlusion) collectively strengthen model interpretability and trustworthiness.

\section{Conclusion}

Our experiments demonstrate that SkinDualGen significantly enhances the performance of classification and segmentation models by generating high-quality synthetic skin lesion images and masks, effectively addressing the challenges of data scarcity and class imbalance. But limitations exist, including potential biases in synthetic data, which may lead to uneven model performance across populations. Second, occasional geometric distortions in masks require refined prompt engineering, increasing application complexity. Future research could extend SkinDualGen to other medical imaging modalities like CT and MRI, strengthen privacy protection and fairness~\cite{fernandez2023privacy}~and 3D medical image generation, ensuring wider clinical applicability.


%
%
%
\bibliographystyle{splncs04}
\bibliography{myref}

\begin{thebibliography}{10}
\providecommand{\url}[1]{\texttt{#1}}
\providecommand{\urlprefix}{URL }
\providecommand{\doi}[1]{https://doi.org/#1}

\bibitem{abdelhalim2021data}
Abdelhalim, I.S.A., Mohamed, M.F., Mahdy, Y.B.: Data augmentation for skin lesion using self-attention based progressive generative adversarial network. Expert Systems with Applications  \textbf{165},  113922 (2021)

\bibitem{montoya2024mam}
Montoya-del Angel, R., Sam-Millan, K., et~al.: Mam-e: Mammographic synthetic image generation with diffusion models. Sensors  \textbf{24}(7), ~2076 (2024)

\bibitem{baur2021autoencoders}
Baur, C., Denner, S., Wiestler, B., Navab, N., Albarqouni, S.: Autoencoders for unsupervised anomaly segmentation in brain mr images: a comparative study. Medical image analysis  \textbf{69},  101952 (2021)

\bibitem{bozorgpour2023dermosegdiff}
Bozorgpour, A., Sadegheih, Y., Kazerouni, A., et~al.: Dermosegdiff: A boundary-aware segmentation diffusion model for skin lesion delineation. In: International workshop on predictive intelligence in medicine. pp. 146--158. Springer (2023)

\bibitem{chen2018encoder}
Chen, L.C., Zhu, Y., Papandreou, G., Schroff, F., Adam, H.: Encoder-decoder with atrous separable convolution for semantic image segmentation. In: Proceedings of the European conference on computer vision (ECCV). pp. 801--818 (2018)

\bibitem{chlap2021review}
Chlap, P., Min, H., Vandenberg, N., Dowling, J., Holloway, L., Haworth, A.: A review of medical image data augmentation techniques for deep learning applications. Journal of medical imaging and radiation oncology  \textbf{65}(5),  545--563 (2021)

\bibitem{dice1945measures}
Dice, L.R.: Measures of the amount of ecologic association between species. Ecology  \textbf{26}(3),  297--302 (1945)

\bibitem{everingham2015pascal}
Everingham, M., Eslami, S.A., Van~Gool, L., Williams, C.K., Winn, J., Zisserman, A.: The pascal visual object classes challenge: A retrospective. International journal of computer vision  \textbf{111},  98--136 (2015)

\bibitem{fernandez2023privacy}
Fernandez, V., Sanchez, P., Pinaya, W.H.L., Jacenk{\'o}w, G., Tsaftaris, S.A., Cardoso, M.J.: Privacy distillation: Reducing re-identification risk of diffusion models. In: International Conference on Medical Image Computing and Computer-Assisted Intervention. pp. 3--13. Springer (2023)

\bibitem{goodfellow2014generative}
Goodfellow, I.J., Pouget-Abadie, J., Mirza, M., Xu, B., et~al.: Generative adversarial nets. Advances in neural information processing systems  \textbf{27} (2014)

\bibitem{he2016deep}
He, K., Zhang, X., Ren, S., Sun, J.: Deep residual learning for image recognition. In: Proceedings of the IEEE conference on computer vision and pattern recognition. pp. 770--778 (2016)

\bibitem{heusel2017gans}
Heusel, M., Ramsauer, H., Unterthiner, T., Nessler, B., Hochreiter, S.: Gans trained by a two time-scale update rule converge to a local nash equilibrium. Advances in neural information processing systems  \textbf{30} (2017)

\bibitem{ho2020denoising}
Ho, J., Jain, A., Abbeel, P.: Denoising diffusion probabilistic models. Advances in neural information processing systems  \textbf{33},  6840--6851 (2020)

\bibitem{hu2022lora}
Hu, E.J., yelong shen, Wallis, P., Allen-Zhu, Z., Li, Y., Wang, S., et~al.: Lo{RA}: Low-rank adaptation of large language models. In: International Conference on Learning Representations (2022)

\bibitem{huang2017densely}
Huang, G., Liu, Z., Van Der~Maaten, L., Weinberger, K.Q.: Densely connected convolutional networks. In: Proceedings of the IEEE conference on computer vision and pattern recognition. pp. 4700--4708 (2017)

\bibitem{iandola2016squeezenet}
Iandola, F.N., Han, S., Moskewicz, M.W., Ashraf, K., Dally, W.J., Keutzer, K.: Squeezenet: Alexnet-level accuracy with 50x fewer parameters and< 0.5 mb model size. arXiv preprint arXiv:1602.07360  (2016)

\bibitem{innani2023generative}
Innani, S., Dutande, P., Baid, U., Pokuri, V., et~al.: Generative adversarial networks based skin lesion segmentation. Scientific Reports  \textbf{13}(1),  13467 (2023)

\bibitem{kazerouni2023diffusion}
Kazerouni, A., Aghdam, E.K., Heidari, M., Azad, R., Fayyaz, M., Hacihaliloglu, I., et~al.: Diffusion models in medical imaging: A comprehensive survey. Medical image analysis  \textbf{88},  102846 (2023)

\bibitem{kingma2013auto}
Kingma, D.P., Welling, M., et~al.: Auto-encoding variational bayes (2013)

\bibitem{koetzier2024generating}
Koetzier, L.R., Wu, J., Mastrodicasa, D., Lutz, A., et~al.: Generating synthetic data for medical imaging. Radiology  \textbf{312}(3),  e232471 (2024)

\bibitem{ktena2024generative}
Ktena, I., Wiles, O., Albuquerque, I., Rebuffi, S.A., Tanno, R., Roy, A.G., et~al.: Generative models improve fairness of medical classifiers under distribution shifts. Nature Medicine  \textbf{30}(4),  1166--1173 (2024)

\bibitem{liu2021swin}
Liu, Z., Lin, Y., Cao, Y., Hu, H., Wei, Y., Zhang, Z., et~al.: Swin transformer: Hierarchical vision transformer using shifted windows. In: Proceedings of the IEEE/CVF international conference on computer vision. pp. 10012--10022 (2021)

\bibitem{medghalchi2024meddap}
Medghalchi, Y., Zakariaei, N., Rahmim, A., Hacihaliloglu, I.: Meddap: Medical dataset enhancement via diversified augmentation pipeline. arXiv preprint arXiv:2403.16335  (2024)

\bibitem{rombach2022high}
Rombach, R., Blattmann, A., Lorenz, D., Esser, P., Ommer, B.: High-resolution image synthesis with latent diffusion models. In: Proceedings of the IEEE/CVF conference on computer vision and pattern recognition. pp. 10684--10695 (2022)

\bibitem{ronneberger2015u}
Ronneberger, O., Fischer, P., Brox, T.: U-net: Convolutional networks for biomedical image segmentation. In: Medical image computing and computer-assisted intervention--MICCAI 2015: 18th international conference, Munich, Germany, October 5-9, 2015, proceedings, part III 18. pp. 234--241. Springer (2015)

\bibitem{ruiz2023dreambooth}
Ruiz, N., Li, Y., Jampani, V., Pritch, Y., Rubinstein, M., Aberman, K.: Dreambooth: Fine tuning text-to-image diffusion models for subject-driven generation. In: Proceedings of the IEEE/CVF conference on computer vision and pattern recognition. pp. 22500--22510 (2023)

\bibitem{shorten2019survey}
Shorten, C., Khoshgoftaar, T.M.: A survey on image data augmentation for deep learning. Journal of big data  \textbf{6}(1),  1--48 (2019)

\bibitem{tschandl2018ham10000}
Tschandl, P., Rosendahl, C., Kittler, H.: The ham10000 dataset, a large collection of multi-source dermatoscopic images of common pigmented skin lesions. Scientific data  \textbf{5}(1), ~1--9 (2018)

\bibitem{yu2024meddiff}
Yu, Y., Gu, Y., Zhang, S., et~al.: Meddiff-fm: A diffusion-based foundation model for versatile medical image applications. arXiv preprint arXiv:2410.15432  (2024)

\bibitem{zhang2023adding}
Zhang, L., Rao, A., Agrawala, M.: Adding conditional control to text-to-image diffusion models. In: Proceedings of the IEEE/CVF international conference on computer vision. pp. 3836--3847 (2023)

\end{thebibliography}
%
\end{document}